\documentclass{epl}
\usepackage{graphicx,psfrag} 
\newcommand{\sign}{\mathop{\mathrm{sign}}} 
 
\title{Quasispecies evolution in general mean-field landscapes} 
\author{Luca Peliti\inst{1}$^,$\inst{2}\footnote{%
Associato INFN, Sezione di Napoli. E-mail: \texttt{peliti@na.infn.it}}} 
\institute{ 
     \inst{1} Institute for Theoretical Physics, University of 
California, Santa Barbara CA 93106--4030 (USA)\\ 
     \inst{2} Dipartimento 
di Scienze Fisiche and Unit\`a INFM, Universit\`a ``Federico II'', 
Complesso Monte S. Angelo, I--80126 Napoli 
(Italy)\footnote{Permanent address.}} 
 
\begin{document} 
 
\maketitle 
 
\begin{abstract} 
I consider a class of fitness landscapes, in which the fitness is a 
function of a finite number of phenotypic ``traits'', which are themselves 
linear functions of the genotype. I show that the stationary trait 
distribution in such a landscape can be explicitly 
evaluated in a suitably defined ``thermodynamic limit'', which is a combination 
of infinite-genome and strong selection limits. These considerations can be 
applied in particular to identify relevant features of the evolution of 
promoter binding sites, in spite of the shortness of the corresponding sequences. 
\end{abstract} 
 
The quasispecies (QS) model~\cite{Eige71,EMS} is extremely useful 
to investigate the behavior of populations evolving in a given fitness 
landscape~\cite{Baake}, although it is based on a rather unrealistic infinite-population 
approximation. It leads to the QS equation, which is a deterministic 
evolution equation for the fraction 
of individuals in the population carrying a given genotype. The dimensionality of 
the QS equation is in principle equal to the number 
of possible genotypes---an enormously large number even for the smallest 
organism. Most analytical treatments of the QS equation have therefore 
focused on situations where this number could be reduced by 
lumping together genotypes in a small number of classes. In some 
``master-sequence'' landscapes, where fitness 
depends only on the Hamming distance from a given, optimal 
genotype~\cite{EMS,Tara92}, it is possible to treat 
together all genotypes whose Hamming distance from the 
master sequence is the same, forming what is called 
an error class. The QS equation can be projected on error classes, 
yielding a vast dimensionality reduction. 
 
However this simplification is not warranted in a number of interesting cases. 
Even in master sequence landscapes, the fitness 
can be a function not only of the number of mutations away from the peak, but 
also of \textit{where} they appear: some sites in the sequence can be more 
important than others. On the other hand, there can be more than one fitness 
peak, quite unrelated to one another. 
 
There have recently been a number of attempts to describe 
evolution in a low-dimensional space of quantitative traits, in a 
so-called \textit{phenotypic} approach~\cite{pheno}. In this case the 
QS equation is low-dimensional from the outset, but the mutation model 
is more or less arbitrary. In particular, one loses the fact that different 
phenotypes can be expressed by greatly varying numbers of genotypes, 
and that equilibrium may be reached from a balance between fitness and 
mutational load. 
 
In the present letter I show how the gap between the genotypic approach 
of the original QS model and the phenomenological phenotypic approach 
can be bridged for a class of fitness landscapes that I shall call the 
\textit{general} mean-field landscapes. The main assumption is that the 
fitness is a function of a finite number of ``phenotypical traits'', which 
are themselves \textit{linear} functions of the genotypical sequence. 
However, the dependence of the fitness on the traits is more or 
less arbitrary. Special cases of these landscapes are the single-peak 
landscape, the Hopfield~\cite{Hopfield,AGS} landscape considered by 
Leuth\"ausser~\cite{Leuth}, and the ``mesa'' landscape 
introduced by Gerland and Hwa~\cite{GH} to model the evolution 
of DNA binding motifs. I shall describe the evolutionary dynamics in 
the ``thermodynamic limit'' introduced in refs.~\cite{FPS,FP}, 
which is close in spirit to the strong selection limit considered by 
Krug~\cite{Krug} to treat the transient in quasispecies evolution 
as a form of extremal dynamics. 
 
I shall discuss my approach within a simple 
two-letter alphabet representation for the genotype. I defer to a further 
publication the generalization to the four-letter alphabet 
and a discussion of a number of more realistic fitness landscapes. 
I first define the QS equation and the general mean-field fitness 
landscapes. I then show how the solution of the QS equation can be reduced 
to an extremal problem within the thermodynamic limit and via an 
additional slow change assumption. I then discuss the consequences 
of this approach in some interesting cases describing interesting 
evolutionary phenomena. The validity of my approximations is briefly 
discussed at the end. 
 
I consider a very large population of individuals evolving according 
to a Darwinian (re\-pro\-duc\-tion-mu\-ta\-tion-se\-lec\-tion) mechanism, 
with a one-parent (asexual) reproduction and with a simple 
point-mutation model of nucleotide substitution. I assume that the 
``genotype'' is described by sequences of $L$ binary units, $\sigma=(\sigma_i)$, 
$i=1,\ldots,L$, $\sigma_i=\pm 1$. These sequences may describe, e.g., a short 
segment of the genome, corresponding to a binding motif, as in~\cite{GH}. 
I also assume non-overlapping generations, and define the fitness 
weight $W_\sigma \ge 0$  to be proportional 
to the expected number of offspring of an individual carrying the genotype $\sigma$. 
In the infinite-population limit, the fraction $x_\sigma(t)$ of individuals 
carrying the genotype $\sigma$ at generation $t$ obeys the QS equation 
\begin{equation}\label{QS:eq} 
    x_\sigma(t+1)=\frac{1}{\left<W\right>_t}\sum_{\sigma'} 
    Q_{\sigma\sigma'}\,W_{\sigma'}\,x_{\sigma'}(t), 
\end{equation} 
where $\left<W\right>_t=\sum_\sigma W_\sigma x_\sigma(t)$ is the average 
fitness weight and $\mathsf{Q}=\left(Q_{\sigma\sigma'}\right)$ is the mutation 
matrix. Within a simple independent mutation model with mutation probability 
per generation equal to $\mu_i$ at unit $i$ one has~\cite{Leuth} 
\begin{equation}\label{mutation:eq} 
  Q_{\sigma\sigma'}=\prod_i\left(\mu_i^{\delta_{\sigma_i\sigma'_i}} 
  (1-\mu_i)^{1-\delta_{\sigma_i\sigma'_i}}\right) 
  \propto  \exp\left(\sum_{i=1}^L\beta_i\sigma_i\sigma'_i\right), 
\end{equation} 
where $\beta_i  =  \frac{1}{2}\log \left(\mu_i^{-1}-1\right)$. 
 
In a general mean-field landscape, the fitness weight $W_\sigma$ depends 
on the genotype $\sigma$ only via a finite number of linear ``traits'' $m^\alpha$, 
$\alpha=1,\ldots,p$, defined by 
\begin{equation}\label{traits:eq} 
  m^\alpha_\sigma=\frac{1}{L}\sum_{i=1}^L \xi^\alpha_i \sigma_i, 
\end{equation} 
as a function of the vectors $\xi^\alpha=(\xi^\alpha_i)$. One can assume 
in the following either that these vectors are known, or that their 
components are independent random variables. Our results will be 
expressed as averages over the distribution of the components of 
the $\xi^\alpha$ in both cases. 
 
The fitness weight $W_\sigma$ then assumes the form 
\begin{equation}\label{fitness:eq} 
  W_\sigma \propto \exp\left(L \,\kappa \phi(\vec m_\sigma)\right), 
\end{equation} 
where $\kappa$ is a selection 
coefficient, $\vec m_\sigma=(m^\alpha_\sigma)$ and $\phi({}\cdot{})$ is a rather 
arbitrary function of its argument. Special cases include: 
\begin{description} 
\item[The master-sequence landscape:] 
Here $p=1$ ($m$ is a scalar) and $\phi(m)$ is, say, a monotonically increasing 
function of $m$. The master sequence corresponds to $\sigma_i=\sign \xi_i^1$. 
The usual sharp-peak landscape corresponds to $\xi_i^\alpha=\pm 1$ and 
$\phi(m)$ equal to 1 if $m=1$ and to 0 otherwise. One can consider in general 
$\phi(m)=m^\lambda$, where $\lambda$ is an epistasis parameter (no epistasis for 
$\lambda=1$, positive epistasis for $\lambda>1$, etc.). The ``mesa'' landscape~\cite{GH} 
corresponds to 
$\phi(m)=\left(1+\exp\left(-\lambda(m-m_0)\right)\right)^{-1}$~\cite{Anderson,APS}, 
where $\lambda$ can be taken to infinity. 
\item[The Hopfield landscape:] 
Here $p > 1$ (but finite), and $\phi(\vec m)$ is a function of the $p$-di\-men\-sion\-al 
vector $\vec m=(m^\alpha)$. In the original Hopfield model one has 
$\phi(\vec m)=\frac12\sum_\alpha \left(m^\alpha\right)^2$, but $\phi$ can be more general. 
However, even if $\phi(\vec m)=\sum_\alpha \phi_\alpha(m^\alpha)$, 
the different components of $\vec m$ are not independent, since adaptation in one 
component may disrupt adaptation in the other. For example, one may consider 
a sequence which should exhibit an affinity larger than some threshold for a given 
factor, and \textit{lower} than another threshold for another one, as in the 
``molecular ecology'' experiments of Ordoukhanian and Joyce~\cite{OJ}, 
in which a Class~I Ligase ribozyme is made to evolve in the presence of a 10--23 
DNA enzyme which binds to the same subsequence as the substrate. 
\item[The Royal Road landscape:] This landscape is rather popular in the 
theory of Genetic Algorithms, because it embodies neutrality and adaptation 
in a simple way~\cite{MHF}. The genotype of length $L=BK$ 
is divided into $K$ blocks of length $B$ each. For each block 
$\alpha$, $m^\alpha$ is defined by $m^\alpha=B^{-1}\sum_{i \in B_\alpha} 
\xi^\alpha_i\sigma_i$, where $B_\alpha$ is the set of units which 
belong to block $\alpha$. The difference with the Hopfield 
landscapes is that the blocks do not overlap. If $\phi$ is 
additive with respect to the blocks, the evolution of each 
block is independent of the other in the quasispecies model. 
The most interesting results are therefore obtained when 
there is epistatic interaction between the different blocks. 
\end{description} 
 
The solution of the QS equation can be expressed as a ``functional 
integral''. One defines $x_\sigma(t)=y_\sigma(t)/\sum_{\sigma'}y_{\sigma'}(t)$, 
where the $y_\sigma$'s follow the \textit{linear} QS equation: 
\begin{equation}\label{linear:eq} 
    y_\sigma(t+1)=\sum_{\sigma'}Q_{\sigma\sigma'}\,W_{\sigma'}\,y_{\sigma'}(t). 
\end{equation} 
One can then write 
\begin{equation}\label{sum:eq} 
y_\sigma(t+1)  \equiv  y_{\sigma(t+1)} = \sum_{\sigma(t)}\cdots\sum_{\sigma(0)} 
\exp\left\{\sum_{\tau=0}^t\left( 
\sum_{i=1}^N\beta_i\sigma_i(\tau+1)\sigma_i(\tau) 
+L\kappa \phi(\vec m_{\sigma(\tau)})\right)\right\} y_{\sigma(0)}. 
\end{equation} 
By using the Fourier representation of the delta function, this expression 
can be written 
\begin{eqnarray} 
&&y_{\sigma(t+1)}\propto \int\prod_{\tau=0}^t 
\left(d \vec m(\tau)\frac{d\vec \lambda(\tau)}{2\pi i}\right) 
\exp\left\{\sum_{\tau=0}^t L 
\left(\kappa \phi(\vec m(\tau))-\vec\lambda(\tau)\cdot\vec m(\tau)\right)\right\} 
\nonumber\\ 
&&\quad{}\times\prod_{i=1}^L\left\{\sum_{\sigma_i(t)}\cdots\sum_{\sigma_i(0)} 
\exp\left[\sum_{\tau=0}^t\sum_{i=1}^L\left(\beta_i\sigma_i(\tau+1)\sigma_i(\tau) 
+\vec \lambda(\tau)\cdot \vec\xi_i\sigma_i(\tau)\right)\right]\right\}y_{\sigma(0)}. 
\end{eqnarray} 
Here we have set, e.g., $\vec \lambda \cdot \vec \xi_i 
=\sum_\alpha \lambda^\alpha\xi^\alpha_i$. 
If we neglect multiple-spin correlation in the initial condition, the second 
line can be written 
\begin{equation} 
\mathcal{J}_{\sigma(t+1)}=\prod_{i=1}^L\sum_{\sigma_i(0)} 
\left\{\prod_{\tau=0}^t\left[ 
\mathsf{K}(\beta_I,\vec\lambda(\tau)\cdot\vec\xi_i)\right]_{\sigma_i(t+1) 
\sigma_i(0)}y_{\sigma_i(0)}\right\}, 
\end{equation} 
where $\mathsf{K}(\beta,h)=\left(\exp(\beta \sigma\sigma'+h\sigma')\right)$ 
is the transfer matrix of a 1D Ising model. 
 
If $\lambda(\tau)$ and $m(\tau)$ are ``slowly varying'' 
one has 
\begin{equation} 
\prod_{\tau=t'}^{t}\mathsf{K}(\beta_i,\vec\lambda(\tau)\cdot\vec\xi_i)\simeq 
\mathsf{K}^{t-t'}(\beta_i,\vec\lambda(\tau)\cdot\vec\xi_i)\simeq 
K^{t-t'}_{\max} (\beta,\vec\lambda(\tau)\cdot\vec\xi_i) 
\mathsf{P}_{\max}(\beta_i,\vec\lambda(\tau)\cdot\vec\xi_i), 
\end{equation} 
where $K_{\max}(\beta_i,h)$ is the larger eigenvalue of $\mathsf{K}(\beta,h)$, 
and $\mathsf{P}_{\max}(\beta,h)$ the projector on the corresponding eigenvector. 
 
Define 
\begin{equation} 
\sum_\sigma \delta\left(L \vec m-L \vec m_\sigma\right) \,y_\sigma(t)= 
\exp\left(L F(\vec m,t)\right). 
\end{equation} 
Then, if in the initial condition $y_\sigma$ depends on $\sigma$ only via $\vec m$, and 
within the slow change approximation, 
\begin{eqnarray}\label{funcint:eq} 
&&\exp\left(L F(\vec m,t+1)\right)  \propto 
\int\prod_{\tau=0}^t d\vec m(\tau)\int\prod_{\tau=0}^{t+1}\frac{d\vec\lambda(\tau)}{2\pi i} 
\exp\left\{\sum_{\tau=0}^{t+1} L \left( 
\kappa \phi(\vec m(\tau))-\vec\lambda(\tau)\cdot\vec m(\tau)\right)\right\}\nonumber\\ 
&&\qquad\times\exp\left\{L\sum_{\tau=0}^t 
\overline{\ln K_{\max}(\beta,\vec\lambda(\tau)\cdot\vec\xi)}+L F(\vec m(0),0)\right\}. 
\end{eqnarray} 
Here $\overline{\ln K_{\max}(\beta,\vec\lambda\cdot\vec\xi)} 
=\mathcal{F}(\vec\lambda(\tau))$ is the average 
of $\ln K_{\max}(\beta,\vec\lambda\cdot\vec\xi)$ 
with respect to the distribution of the $\beta$'s and of 
the $\vec\xi$'s. This can be either the average over the actual, 
known distribution of these quantities, if one is lucky enough 
to know it; or over some reasonable \textit{a priori} distribution 
otherwise. 
 
\begin{figure} 
\psfrag{m}[b]{$m$} 
\psfrag{m0}[l]{$m$}
\psfrag{Gamma}[b]{$\kappa \phi+\Gamma$} 
\psfrag{betaL2to3}[l]{$(\beta-\beta_\mathrm{th})L^{2/3}$} 
\begin{center} 
\begin{tabular}{ccc} 
\includegraphics[width=6cm]{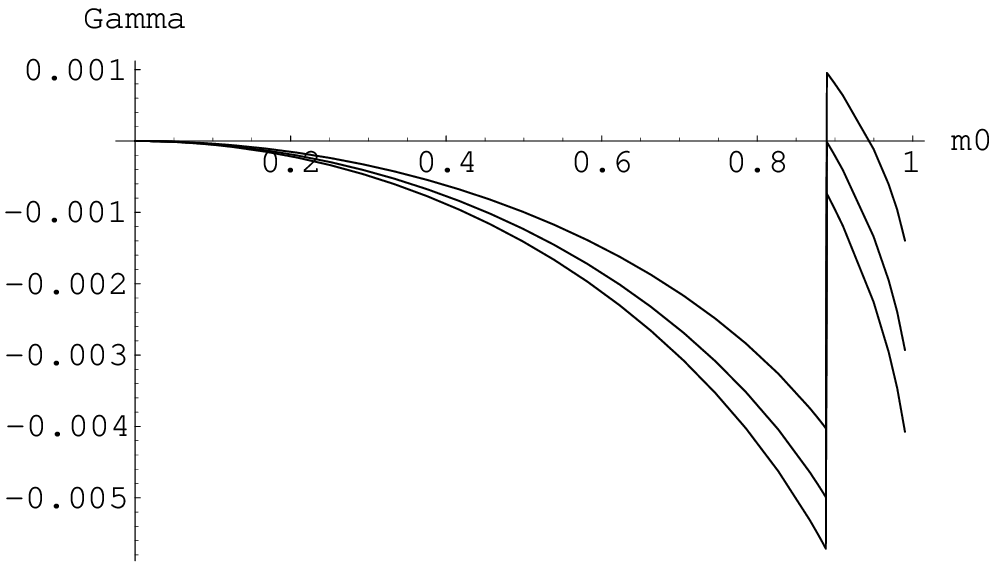} 
&\hspace{0.2cm}&
\includegraphics[width=6cm]{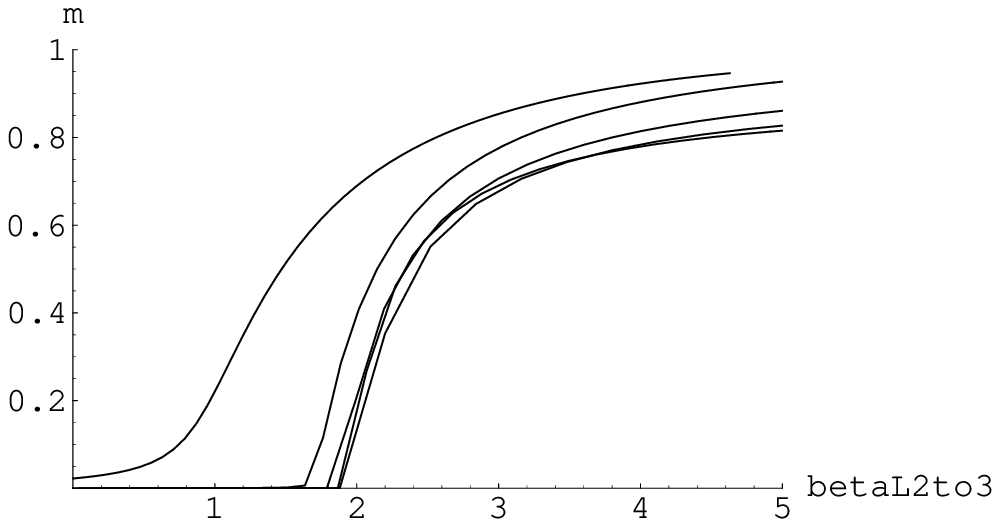} 
\end{tabular} 
\end{center} 
\caption{Left: The error threshold for the $p=1$ case in the mesa landscape 
$\phi(m)=\theta(m-m_0)$. The function $\kappa \phi(m) 
+\Gamma(m)$ is plotted vs.\ $m$ for (from bottom up) $\beta=2.25,2.3425,2.45$. 
We have set $\kappa=0.005$ and $m_0=8/9$. 
One observes that the maximum moves from the origin to 
$m_0$ for $\beta=\beta_\mathrm{th}=2.3425$. Right: Finite-size 
scaling $m=m(\beta,L)$ for $\kappa=0.005$ and $m_0=8/9$. 
X axis: $(\beta-\beta_\mathrm{th})L^{2/3}$ with $\beta_\mathrm{th}=2.3425$. 
From left to right: $L=8,16,32,64,128$.} 
\label{mesa:fig} 
\end{figure}

If $L$ is large enough, this expression can be evaluated by the saddle point 
method. In particular, the error threshold can be identified at 
stationarity, by looking at the extremum of the function 
$\kappa \phi(\vec m)-\vec\lambda\cdot\vec m+\mathcal{F}(\vec\lambda)$ 
with respect to $(\vec m,\vec \lambda)$. 
This corresponds to the extremum $\vec m^*$ 
of $\kappa\phi(\vec m)+\Gamma(\vec m)$, where $\Gamma(\vec m)$ is the 
Legendre transform of $\mathcal{F}(\vec \lambda)$ with respect to 
$\vec \lambda$.\footnote{This value is not equal to the \textit{actual} 
average value of $m$ in the stationary population \cite{Tara92}, 
but is close enough to it if the mutation rate is small.} 
If the distribution of the $\xi$'s is symmetrical, the 
maximum of $\Gamma(m)$ is located at the origin. As $\beta$ 
increases (i.e., as the mutation rate gets smaller), $\Gamma(m)$ 
becomes flatter and flatter. As $\beta$ becomes larger than a threshold 
value $\beta_\mathrm{th}$, $m^*$ moves away from the origin: this 
is the error threshold: see fig.~\ref{mesa:fig}~(left). 
Within a simple ``mesa'' landscape, the error threshold 
is approximately located at the point in which $\kappa\phi(m_0)+\Gamma(m_0)$ 
becomes larger than $\Gamma(0)$. For $\beta > \beta_\mathrm{th}$, 
the optimum $m^*$ remains close to $m_0$, except (for finite $\lambda$) 
at extremely small mutation rates, in agreement with the results 
of ref.~\cite{GH,Goldstein}. 

In order to analyze the finite-length behavior
of the system, it is useful to apply the Schr\"odinger equation
approach to the quasispecies equation~\cite{BBW1,GH}. The
role of the quantum constant $\hbar$ is played by
the inverse of the genome length $L$. In the simple mesa landscape,
the quasispecies equation can be transformed into a Schr\"odinger
equation in a potential which is the superposition of a harmonic
oscillator potential near $m=0$ and a linear potential
with a barrier near the threshold. The finite-length
threshold $\beta_\mathrm{th}(L)$ can be identified by the
condition that the ground-state energies near the two classical
minima are equal. One thus obtains the result that  $\beta_\mathrm{th}(L)$
reaches its asymptotic value as
$\beta_\mathrm{th}(L)=\beta_\mathrm{th}(\infty)+\mathrm{O}(L^{-2/3})$, 
rather than $\mathrm{O}(L^{-1})$, which holds
for the sharp-peak landscape (cf.\ ref.~\cite{Tara92}) or for smoother
ones. The width of the distribution in $m$ above the threshold also
behaves like $L^{-2/3}$. The transition is 
however quite sharp even for moderate values of $L$, as can
be seen by the finite-scaling analysis shown in fig.~\ref{mesa:fig}~(right).
 
\begin{figure}[htb] 
\psfrag{beta}[l]{$\beta$} \psfrag{kappa2}{$\kappa_2$} 
\begin{center} 
\includegraphics[height=4cm]{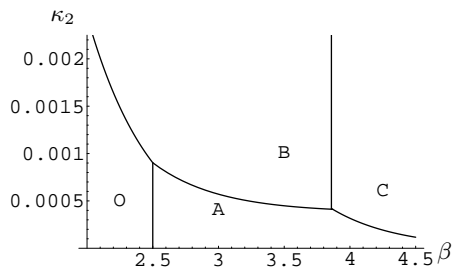} 
\end{center} 
\caption{The phase diagram in the $(\beta,\kappa_2)$ plane for the $p{=}2$ 
Hopfield landscape, with $\kappa\phi(\vec m)=\kappa_1\theta(m^1-0.33)+\kappa_2 
\theta(m^2-0.5)$, and $\kappa_1=0.377417\,10^{-4}$. 
The letters denote the stability regions for the points O: $\vec m^*=(0,0)$, 
A: $\vec m^*=(0.33,0)$, B: $\vec m^*=(0,0.5)$, and C: $\vec m^*=(0.33,0.5)$.} 
\label{Hopfield:fig} 
\end{figure}

I discuss the Hopfield landscape in the didactically simple case 
of $p=2$, $\xi^\alpha_i=\pm 1$. One can easily evaluate $\Gamma(\vec m)$ 
numerically: it exhibits a maximum at $\vec m=0$, and is higher 
on the axes. Again, it becomes flatter and flatter as $\beta$ increases. 
Let us consider the case in which $\kappa\phi(\vec m)=\kappa_1\theta(m^1-m^1_0) 
+\kappa_2\theta(m^2-m^2_0)$. One can identify the error thresholds 
$(\beta^1_\mathrm{th},\beta^2_\mathrm{th})$ on 
the two axes, and the actual threshold will take place at the smaller 
$\beta_\mathrm{th}$. However, there might be a second threshold at 
a larger value of $\beta$, where $\vec m^*$ moves from one axis to 
another, and even a third one, where $\vec m^*$ acquires two nonzero 
components. See fig.~\ref{Hopfield:fig}. 
 
In the Royal Road case, each block will have its own $\Gamma(m)$ 
function. If the fitness function $\phi(\vec m)$ is a sum of contributions, 
one for each block, the stationary point will be determined independently 
for each component as the optimum of 
$\kappa_\alpha \phi_\alpha(m^\alpha)+\Gamma(m^\alpha)$. On the other hand, 
if there is epistatic interaction among blocks, appearing in $\phi(\vec m)$, 
one can find a more complex phase diagram. 
 
Summarizing, I have shown how it is possible to solve for the stationary 
behavior of the quasispecies equation in a number of nontrivial fitness landscapes, 
provided the ``thermodynamic'' and the slow change limits are taken. 
The thermodynamic limit seems far-fetched if one is considering, as in \cite{GH}, 
the evolution of binding motifs. Nevertheless, the error threshold is well 
identified by the present approach, and the basic conclusion that $m^*$ is 
close to the threshold follows directly. The transition appears to be first-order 
in our language, since it corresponds to the ``bulk'' transition, while in \cite{GH} 
it is described as the corresponding wetting transition near a wall (cf.~\cite{Tara92}). 
The slow change limit applies in the stationary regime, and can also be valid in 
the transient regime if the mutation rates are not too small: the condition is that 
the number of generations needed to equilibrate with a given value of $\lambda$ 
should be smaller than the number of generations in which $\lambda$ itself varies 
significantly. This is true if the selective pressures are not too large, and the 
mutation rates not too small. The application of the present approach to the dynamics 
is a problem worth further investigation.

\acknowledgments 
This research was supported in part by the National 
Science Foundation under Grant No.~PHY99-07949, and 
has been performed within a joint cooperation 
agreement between Japan Science and Technology Corporation (JST) and 
Universit\`a di Napoli \lq\lq Federico II\rq\rq. Interesting conversations 
and suggestions by U. Gerland, T. Hwa and J. Krug are 
gratefully acknowledged. 
This work is dedicated to Raffaele (Peo) Tecce, in fond 
remembrance of fascinating conversations on the architecture 
of life.

\end{document}